\documentclass[twocolumn,prl,showpacs,superscriptaddress,amssymb,amsmath,amsmath,floatfix]{revtex4-1}
\usepackage{graphicx}
\usepackage{epstopdf}
\usepackage{bm}
\usepackage[colorlinks=true,linkcolor=blue,urlcolor=blue,citecolor=blue]{hyperref}
\usepackage{comment}
\usepackage{float}
\usepackage{cancel}
\usepackage{times}
\usepackage{xcolor}

\newcommand{\be}{\begin{equation}}
\newcommand{\ee}{\end{equation}}

\newcommand{\affFUW}{Faculty of Physics, University of Warsaw, Pasteura 5, 02-093 Warsaw, Poland}
\newcommand{\affICHF}{Institute of Physical Chemistry, Polish Academy of Sciences, Kasprzaka 44/52, 01-224 Warsaw, Poland}
\begin{document}
\title{Collisional losses of ultracold molecules due to intermediate complex formation}
\author{Krzysztof Jachymski}
\email{krzysztof.jachymski@fuw.edu.pl}
\affiliation{\affFUW}
\author{Marcin Gronowski}
\affiliation{\affFUW}\affiliation{\affICHF}
\author{Micha{\l} Tomza} 
\email{michal.tomza@fuw.edu.pl}
\affiliation{\affFUW}

\date{\today}

\begin{abstract}

Understanding the sources of losses and chemical reactions of ultracold alkali-metal molecules is among the critical elements needed for their application in precision measurements and quantum technologies. Recent experiments with nonreactive systems have reported unexpectedly large loss rates, posing a challenge for theoretical explanation. Here, we examine the dynamics of intermediate four-atom complexes formed in bimolecular collisions. We calculate the nuclear spin--rotation,  spin--spin, and quadrupole coupling constants for bialkali four-atom complexes using \textit{ab intio} quantum-chemical methods. We show that the nuclear spin--spin and quadrupole couplings are strong enough to couple different rotational manifolds to increase the density of states and lifetimes of the collision complexes, which is consistent with experimental results. We also reveal that the interaction-induced variation of electron spin--nuclear spin couplings explains the recently observed long lifetime of alkali-metal three-atom complexes formed in atom-molecule collisions. Finall, we propose further experiments to confirm our predictions.

\end{abstract}

\maketitle

{\it Introduction.}
Creating ultracold dense samples of polar molecules has been a long-standing goal that proved to be much more challenging to realize than producing quantum degenerate atomic gases~\cite{Quemener2012}. Nevertheless, it is worth pursuing due to their potential for quantum simulations, quantum computing, controlled chemical reactions studies, and precision measurements that take advantage of the rich molecular structure and strong intermolecular dipolar interactions~\cite{CarrNJP09,Balakrishnan2016,Bohn2017}. Multiple approaches towards the quantum collisional regime have been studied, including crossed and merged beam techniques~\cite{Gilijamse2006,Henson2012,Jankunas2015}, laser cooling~\cite{ShumanNature10,TruppeNP17,Anderegg2018}, as well as the association of molecules from already ultracold atoms~\cite{Ni2008,Takekoshi2014,Molony2014,ParkPRL15,Guo2016,VogesPRL20}. There has been remarkable experimental progress along the lines mentioned above, allowing for studies of collisional dynamics in the ultracold regime, demonstrating the effects of particle statistics, initial quantum state, and long-range interactions~\cite{Ni2010,Ospelkaus2010,MirandaNatPhys11,YeSA18,DeMarcoScience19,HuScience19,GregoryNC19,Liu2020,HuNC21}, among other breakthrough results.

While ultracold inelastic collisions and chemical reactions represent a fascinating area of study, it is necessary to learn how to mitigate them to prolong the system lifetime and reach quantum degeneracy by evaporative cooling~\cite{ValtolinaNature20}. Certain molecular species such as KRb~\cite{Ni2008} are highly reactive~\cite{Ospelkaus2010,HuScience19}, whereas other molecular gases such as RbCs~\cite{Takekoshi2014,Molony2014}, NaRb~\cite{ParkPRL15,VogesPRL20}, and NaK~\cite{Guo2016} should be chemically stable with respect to molecule-molecule collisions~\cite{Zuchowski2010}, leading to longer lifetimes of those systems. Various scenarios of suppressing inelastic collisions by external confinement~\cite{Ni2010,Micheli2010} and microwave or electric field shielding~\cite{Karman2018,LassablierePRL18,Matsuda2020,XiePRL20,Li2021,AndereggScience21} have been proposed and tested, but they further increase the complexity of the experiment. Finding suitable molecular species with strong interactions but very low inelastic and reactive rate constants is thus vital for further progress.

First experiments with ultracold RbCs molecules revealed a somewhat surprising result, as strong two-body losses have been observed even though the chemical reaction in the ground rovibrational state is energetically forbidden in this system~\cite{GregoryNC19}. This could be explained by the sticking mechanism in which the collision creates a long-lived intermediate four-atom complex which is then lost due to further inelastic collisions~\cite{MaylePRA2013,CroftPRA14}. Such processes involving a weakly bound tetratomic state and large available phase space can currently only be treated with simplified methods~\cite{KlosSR21}. However, it has been argued on the grounds of semiclassical statistical models that the complex lifetimes should not be long enough to result in universal (unitarity-limited) losses~\cite{ChristiansenPRA19}. Instead, the complex decay is mainly due to the excitations by the laser light used for optical trapping~\cite{ChristianenPRL19}. This was initially confirmed experimentally~\cite{GregoryPRL20,LiuNP2020}. Recently, new experiments with ultracold NaRb and NaK molecules have rechallenged the theoretical models, reporting significant losses even when the trapping light is switched off periodically~\cite{Bause2021,Gersema2021}, indicating that nuclear spins of the complex may not be conserved, which would increase the density of accessible states and thus also its lifetime.

In this letter, we demonstrate that the couplings between the internal states of ultracold collisional complexes can indeed lead to exploration of the entire phase space of four-atom states and significantly increase the lifetime of the metastable intermediate state. To this end, we calculate the relevant coupling constants for several molecular pairs {\it ab initio}, showing that they compete with other relevant energy scales. This motivates us to extend the conventional rate equations describing the molecular loss in the presence of trapping light and identify the regimes resulting from competing time scales. Our results should be helpful for designing further experiments with ultracold nonreactive molecules.

{\it Molecular collisions.}
We consider a gas of ultracold heteronuclear molecules $AB$, each composed of two alkali-metal atoms. The molecules can be either bosonic or fermionic and are prepared in the lowest electronic, rovibrational, and hyperfine states. The temperature of the gas is assumed to be low enough such that only the first few partial waves matter for scattering properties, but above the quantum degeneracy where collective and quantum coherent effects would need to be considered~\cite{He2020}. Similar conditions are typically encountered in experiments. For simplicity, we assume no external electric field such that intermolecular van der Waals interactions are dominating.

\begin{figure}
\includegraphics[width=\columnwidth]{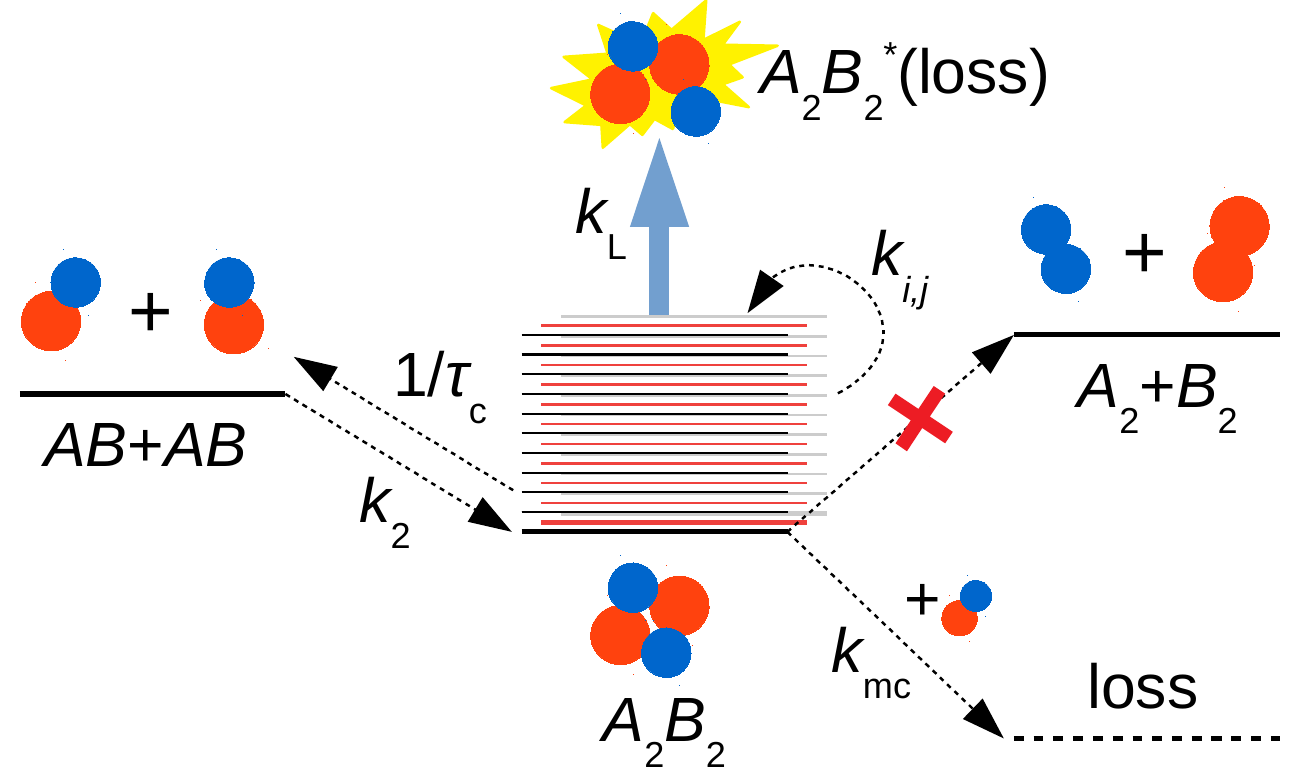}
\caption{Schematic representation of the competing processes in ultracold nonreactive bimolecular $AB+AB$ collisions. Four-atom $A_2B_2$ complexes form with the rate constant $k_2$ and spontaneously dissolve with the rate constant $1/\tau_\text{c}$. They change their internal spin states with the rate constants $k_{ij}$ and can be lost due to the laser photoexcitation with the rate constant $k_\text{L}$ and collisions with other molecules with the rate constant $k_\text{mc}$. In this work, we consider systems where a direct chemical reaction to $A_2+B_2$ is energetically forbidden.}
\label{fig:scheme}
\end{figure}

The energy of the $A_2+B_2$ configuration can be higher or lower than of the initial $AB+AB$ arrangement, opening the possibility for a chemical reaction in the second scenario. In both cases, an intermediate bimolecular complex can be formed during the collision to provide a possible source of additional losses (see Fig.~\ref{fig:scheme}). The other important effect comes from an external optical trapping potential which is commonly present in the experiment. Although the laser is nonresonant with respect to the molecules, it can excite the transient complex to an excited electronic state which will live long enough to always lead to loss~\cite{ChristianenPRL19}. 

Since {\it ab initio} methods are generally too expensive computationally to describe exactly the dynamics of heavy molecule-molecule systems, one has to turn to simplified approaches. The established molecular collision models are based on the known long-range intermolecular interactions, which are critical for scattering at low kinetic energy~\cite{Quemener2012}. The short-range physics can then be described, e.g., in terms of the quantum defect theory~\cite{Mies1984,Croft2011}. Mayle {\it et al.}~\cite{MaylePRA2013} proposed using a statistical short-range boundary condition that includes random couplings to multiple resonant states. The structure of the couplings and resonance positions can be generated based on random matrix theory, which implicitly assumes that the complex dynamics is ergodic. The crucial quantity for this statistical model is the mean complex lifetime $\tau_c$, which can be linked to the density of states $\rho$ and the number of available exit channels $\mathcal{N}$ 
\begin{equation}
	\tau_c=2\pi\hbar\rho/\mathcal{N}\, .
\end{equation}
As $\tau_c$ can easily reach milliseconds, molecular scattering in this regime is often described as ``sticky collisions''. The rate constant for the complex creation has been shown~\cite{MaylePRA2013,Christianen2021} to approach the universal rate constant for low energy inelastic processes based only on the long-range potential~\cite{IdziaszekPRL2010}. For smaller resonance densities, quantum interference effects and shape resonances can significantly modify this behavior~\cite{JachymskiPRL2012}. In this case, the assumption of ergodicity of the complex dynamics can easily break down. Other important effects include the emergence of quantum scars~\cite{Heller1984,Kendrick2021}. It is thus crucial to learn about the internal structure of the complex state in more detail.

{\it Internal complex structure.}
In closed-shell molecules, three main types of internal state couplings can play a role~\cite{PuzzariniIRPC2010}. The nuclear quadrupole coupling (NQC) arises from the interaction of the nuclear quadrupole moments with the electric-field gradient at the corresponding nuclei. The relevant Hamiltonian reads
\begin{equation}
\begin{split}
\hat{H}_\text{NQC}=&\sum_{L} \frac{e Q_L q_J^L}{2 I_L (2I_L-1)J(2J-1) }\left[3(\hat{\bm{I}}_L\cdot\hat{\bm{J}})^2 \right. \\
& + \left.\frac{3}{2} \hat{\bm{I}}_L\cdot\hat{\bm{J}}- \hat{\bm{I}}_L^2\hat{\bm{J}^2}\right]\,,
\end{split}
\label{eq:nqc}
\end{equation}
where the summation is over all nuclei $L$ with a nonzero quadrupole moment, $\hat{\bm{J}}$ and $\hat{\bm{I}}_L$ are the rotation and nuclear spin operators, $J$ and $I_L$ denote the rotation and nuclear spin quantum numbers, $Q_L$ is the nuclear quadrupole moment, and $q_J^L$ is the expectation value of the space-fixed electric-field gradient tensor at the given nucleus averaged over the rotational motion. For linear molecules, the resulting hyperfine splitting energy scale is given by $eQ_L q_L$, where $q_L$ stands for the electric field gradient along the interatomic axis. The nuclear spin--nuclear spin (NSNS) interaction is a sum of direct dipolar and  indirect electron-mediated terms and for linear molecules is given by
\begin{equation}
\hat{H}_\text{NSNS}=\sum_{L \neq M} \left(c_3^{L,M} \hat{\bm{I}}_L\cdot \bm{D_T} \cdot \hat{\bm{I}}_M + c_4^{L,M} \hat{\bm{I}}_L \cdot \hat{\bm{I}}_M\right)\,,
\end{equation}
where $\bm{D_T}$ is the dipolar spin--spin coupling tensor and $c_3^{L,M}$ and $c_4^{L,M}$ are the spin--spin coupling constants. Finally, the nuclear spin--rotation (NSR) coupling arises from the rotation-induced magnetic field at nuclei $L$ described as
\begin{equation}
\hat{H}_\text{NSR}=\sum_L \hat{\bm{I}}_L\cdot\bm{C}_L\cdot\hat{\bm{J}} .
\label{eq:nsr}
\end{equation}
For linear molecules, the tensor $\bm{C}_L$ reduces to the single spin--rotation coupling constant $c_{L}$. Note that here we rely on the rigid rotor approximation, which for a resonant tetratomic state is a rather crude approximation, but it should provide the correct order of magnitude estimates.

The hyperfine parameters for alkali-metal dimers are known~\cite{AldegundePRA2008,AldegundePRA2009,AldegundePRA2017}. Their variation during four-atom complex formation is studied in this work. We present a linear (NaK)$_2$ complex in three configurations as a computational example, but similar results are observed for nonlinear geometries, as well as for (RbCs)$_2$ and (NaRb)$_2$ complexes. We study hyperfine parameters as a function of the distance between centers of mass of the monomers, $R_{c.m.}$. The NQC constants for linear structures of (NaK)$_2$ are computed using the coupled-cluster method including single and double excitations (CCSD)~\cite{BartlettMusial2007} with relativistic effects described by the one-electron variant of the Spin-free Exact Two-component Theory~\cite{ChengGauss2011} as implemented in Cfour~2.1~\cite{CFOUR}. All-electron core-valence relativistic triple-zeta basis sets (aug-cc-pCVTZ-DK for Na and aug-cc-pwCVTZ-X2C for K) are employed~\cite{bas1,bas2}. The NQC constants for nonlinear structures, the NSNS coupling parameters~\cite{VisserJCP1992}, and the NSR coupling constant~\cite{AucarJCP2012} are computed at the Dirac-Hartree-Fock level of theory with the all-electron triple-zeta fully relativistic basis sets~\cite{Dyall2016} as implemented in Dirac 2019~\cite{DIRAC19}.

\begin{figure}
\includegraphics[width=\columnwidth]{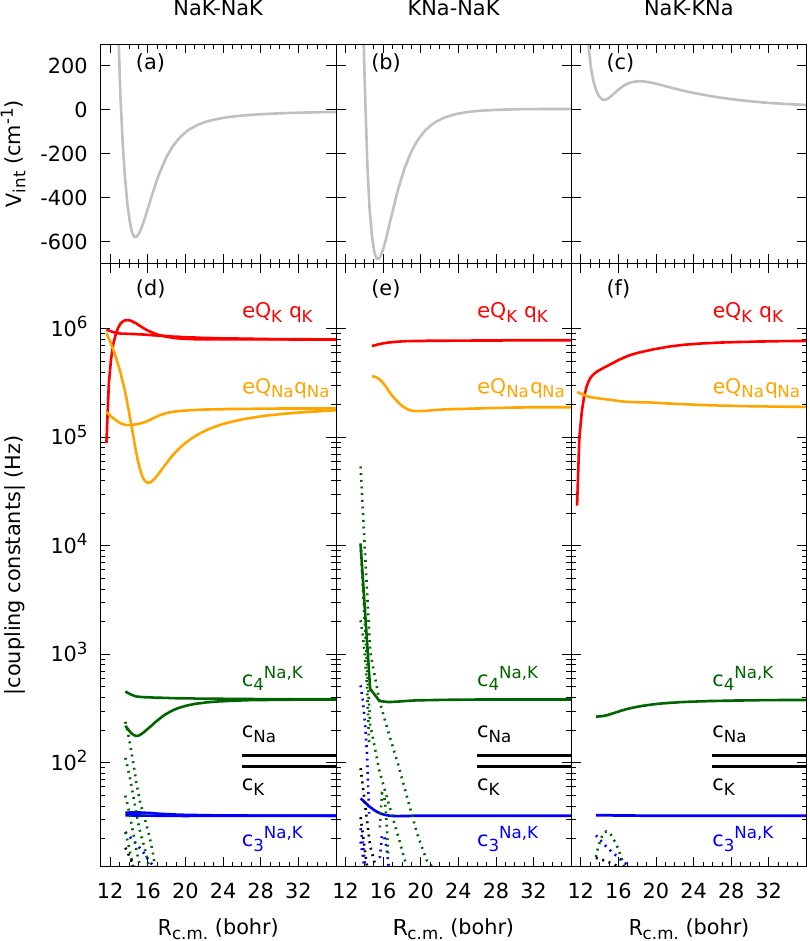}
\caption{Variation of interaction energy (a), (b), (c) and hyperfine parameters (d), (e), (f) of the linear ($^{23}$Na$^{39}$K)$_2$ complex as a function of the intermolecular distance for different intermolecular orientations. The NQC constant on $^{39}$K (red), on $^{23}$Na (orange), nuclear spin--rotation constant (black), nuclear spin--nuclear spin $c_3$ (blue) and $c_4$ (green) are preseted. The solid lines present the hyperfine parameters of diatomic molecules; dotted lines stand for the parameters resulting from intermolecular interactions.}
\label{fig:all}
\end{figure}

Figure~\ref{fig:all} shows the hyperfine parameters as a function of the intermolecular distance for three different linear configurations of the ($^{23}$Na$^{39}$K)$_2$ complex~\cite{SupMat}. When dipole moments of molecules are parallel (NaK-NaK), the complex contains four non-equivalent nuclei. When dipole moments are antiparallel, two centrosymmetric arrangements are possible (KNa-NaK and NaK-KNa). We observe that the magnitude of hyperfine parameters rather weakly depends on the intermolecular distance, as the van der Waals interaction between the molecules is too weak to significantly affect the electronic wave function in the vicinity of the nuclei. New NSNS interactions appear when a complex is formed. We find that NQC contributes the most to the hyperfine structure with interaction-induced variation reaching 1$\,$MHz. While in principle NQC is active only for states with $J>0$, the bimolecular states can involve nonzero rotation even if the total angular momentum vanishes. In addition, one can expect strong mixing of different hyperfine manifolds for $J=0$ in higher orders of perturbation theory, where first the anisotropy of the intermolecular interaction couples to dimers in excited rotational states. The isotropic NSNS interaction is two orders of magnitude weaker than NQC. Still, it is the strongest of all remaining hyperfine interactions and the most important one for states of two nonrotating molecules where $j_1=j_2=J=0$. Interestingly, in the vicinity of the potential energy minimum, the intermolecular NSNS strengthens and can be as large as the intramolecular NSNS ones. Finally, for global minima of the potential energy surface (which is nonlinear \cite{ByrdJCP2012}), the computed isotropic NSNS coupling constant between sodium atoms is $0.5\,$kHz (nearly four times bigger than the isotropic NSNS coupling constant for any pair of Na and K). The crucial result of this calculation is that the internal spin dynamics of the transient collision complex can be governed by energy scales of the order of kHz, competing with other processes happening at larger distances.

{\it Effective collision model.}
We now introduce an effective model for the loss of molecules motivated by the results of the previous section, which extends the usual approach (see Fig.~\ref{fig:scheme}). Specifically, we assume that the collision can lead to creating a transient state conserving the nuclear spin configuration, which can subsequently undergo spin-changing dynamics to other nuclear spin states due to the coupling terms discussed above. All complexes are prone to photoexcitation by the trapping laser regardless of their hyperfine state. We describe the reaction kinetics classically in terms of the molecular density $n_{\rm mol}$, the initially created tetratomic state density $n_{c,0}$, and the other complex state densities $n_{c,i}$ (note that in the presence of the electric fields multiple tetratomic states with different angular momenta can be accessed directly~\cite{Quemener2021}). The rate equations describing the population of different states read
\begin{equation}
\dot{n}_{\rm mol} = -k_2 n_{\rm mol}^2+\frac{2}{\tau_c}n_{c,0}-k_{mc}\sum_j{n_{c,j}}n_{\rm mol}
\end{equation}
\begin{equation}
\begin{split}
\dot{n}_{c,0} =\frac{1}{2} k_2 n_{\rm mol}^2-\frac{1}{\tau_c}n_{c,0}-k_L I(t)n_{c,0}-k_{mc}n_{\rm mol}n_{c,0}\\ + \sum_j{}{\vphantom{\sum}}' k_{0,j}n_{c,j}-\sum_j{}{\vphantom{\sum}}' k_{j,0}n_{c,0}
\end{split}
\end{equation}
\begin{equation}
\dot{n}_{c,i} = -k_L I(t)n_{c,i}-k_{mc}n_{\rm mol}n_{c,i}+ \sum_j{}{\vphantom{\sum}}' k_{i,j}n_{c,j}-\sum_j{}{\vphantom{\sum}}' k_{j,i}n_{c,i}
\end{equation}
Here, $k_2$ is the complex formation rate, $\tau_c$ is its decay time back to a molecular pair (note that other complex states cannot directly decay), $k_{mc}$ is the inelastic molecule-complex collision rate, $k_{i,j}$ are the internal couplings between the complex states, and the primed sums indicate that the state under consideration is excluded. Furthermore, $k_L$ is the photoexcitation rate of the complexes due to light intensity $I(t)$. We neglect complex-complex collisions due to their low overall population which for realistic experimental parameters is typically at least two orders of magnitude lower than molecular density. For simplicity we assume universal rate constant values~\cite{IdziaszekPRL2010} for $k_2$ and $k_{mc}$, which does not fundamentally affect the conclusions. The rate constants $k_{ij}$ responsible for state changing processes are assumed to be random with a certain mean value that can be expressed in units of $1/\tau_c$. While they are fundamentally linked with the nuclear spin couplings, making a direct connection would require solving the multichannel scattering problem, which cannot at the moment be performed with sufficient precision for bimolecular collisions. Compared to the previous approaches, here we separate the initially created metastable state from the rest and assume that only this state can decay back to a pair of molecules. 
% As the internal state coupling strength grows, the loss of molecules becomes faster. This can be understood as a growth of the effective complex lifetime, in agreement with classical ergodic phase space exploration arguments. However, for photoexcitation rates comparable with the other timescales this process can dominate over the spin flips and quickly saturate the sample. This will be shown in detail below.

\begin{figure}
\centering
\includegraphics[width=0.45\textwidth]{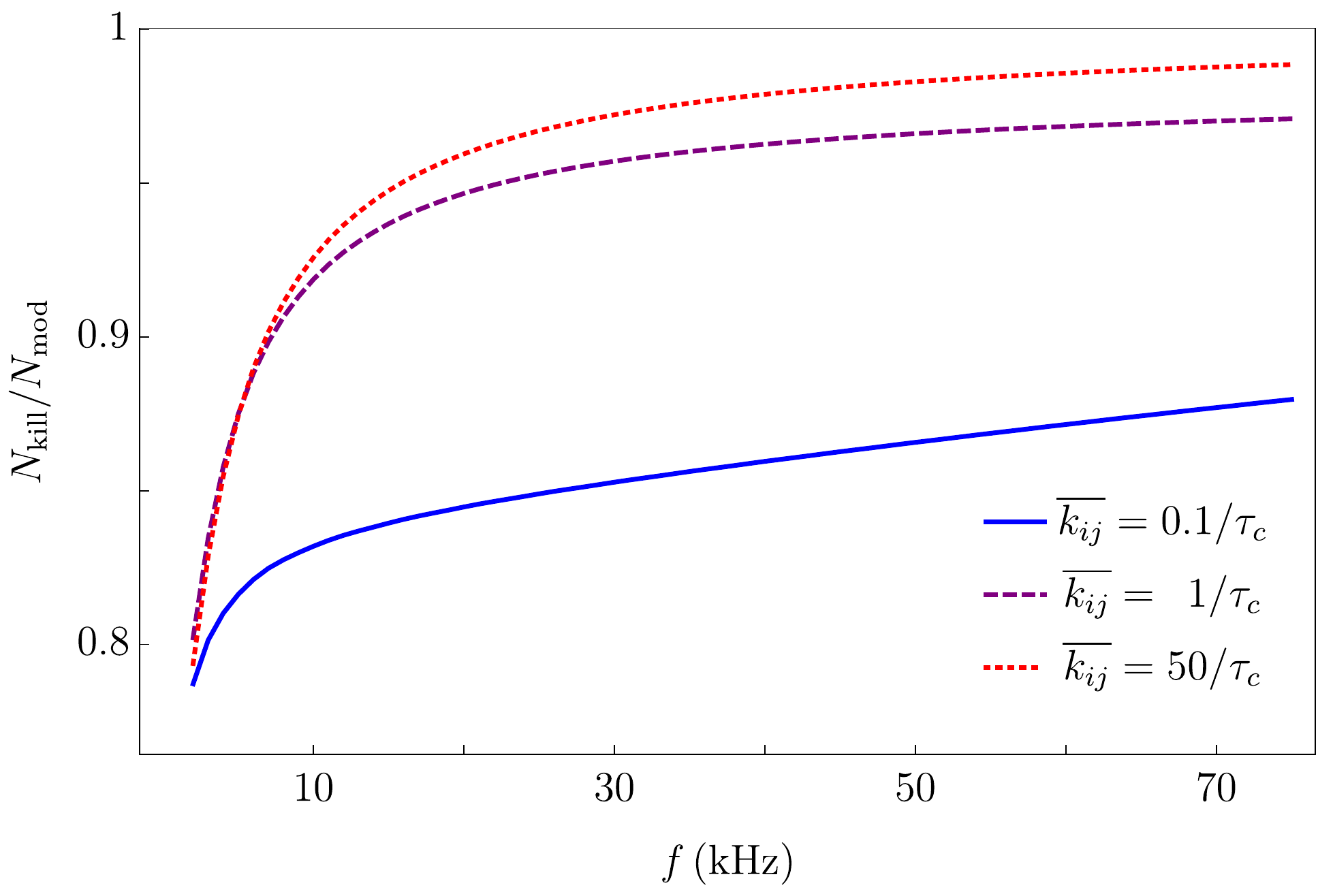}
\caption{\label{fig:nak} The ratio of molecules left in the trap with ($N_\text{kill}$) and without ($N_\text{mod}$) the additional light beam after waiting time of $250\,$ms as a function of the chopping rate $f$ for $^{23}$Na$^{40}$K molecules assuming varying mean internal state coupling strength $\overline{k_{i,j}}$.}
%=0.1/\tau_c$ (blue solid), $1/\tau_c$ (purple dashed), and $50/\tau_c$ (red dotted).}
\end{figure}

{\it Experimental implications.}
Let us now illustrate the predictions of this model and apply them to discuss recent experimental results obtained for several molecular species. The experiments are typically performed as follows: a molecular sample is prepared in the lowest electronic and rovibrational state and then held in a trap that is periodically turned on and off (``chopped'') with various tunable frequency $f$ and operation time. The results are compared with the measurement in the presence of an additional stationary laser light (``kill beam''). One can expect a strong dependence of losses on the relation between the dark time (no trapping light) controlled by the modulation frequency and the complex lifetime, allowing to estimate the latter from experimental data.

First, we briefly consider the experiment performed with $^{40}$K$^{87}$Rb molecules~\cite{HuScience19,LiuNP2020,HuNC21}, which are reactive. In this case, it was possible to experimentally verify that the nuclear spins are conserved during the inelastic process, and the measured complex lifetime turned out to be $\tau_c=360\pm 30\,$ns, while the semiclassical theory estimate is $170\pm 60\,$ns. In this system, the number of open channels is rather large, and presumably, the collision complex almost immediately falls into the region of the phase space corresponding to reaction products, not giving time for spin-changing processes to happen. 

\begin{figure}
\centering
\includegraphics[width=0.45\textwidth]{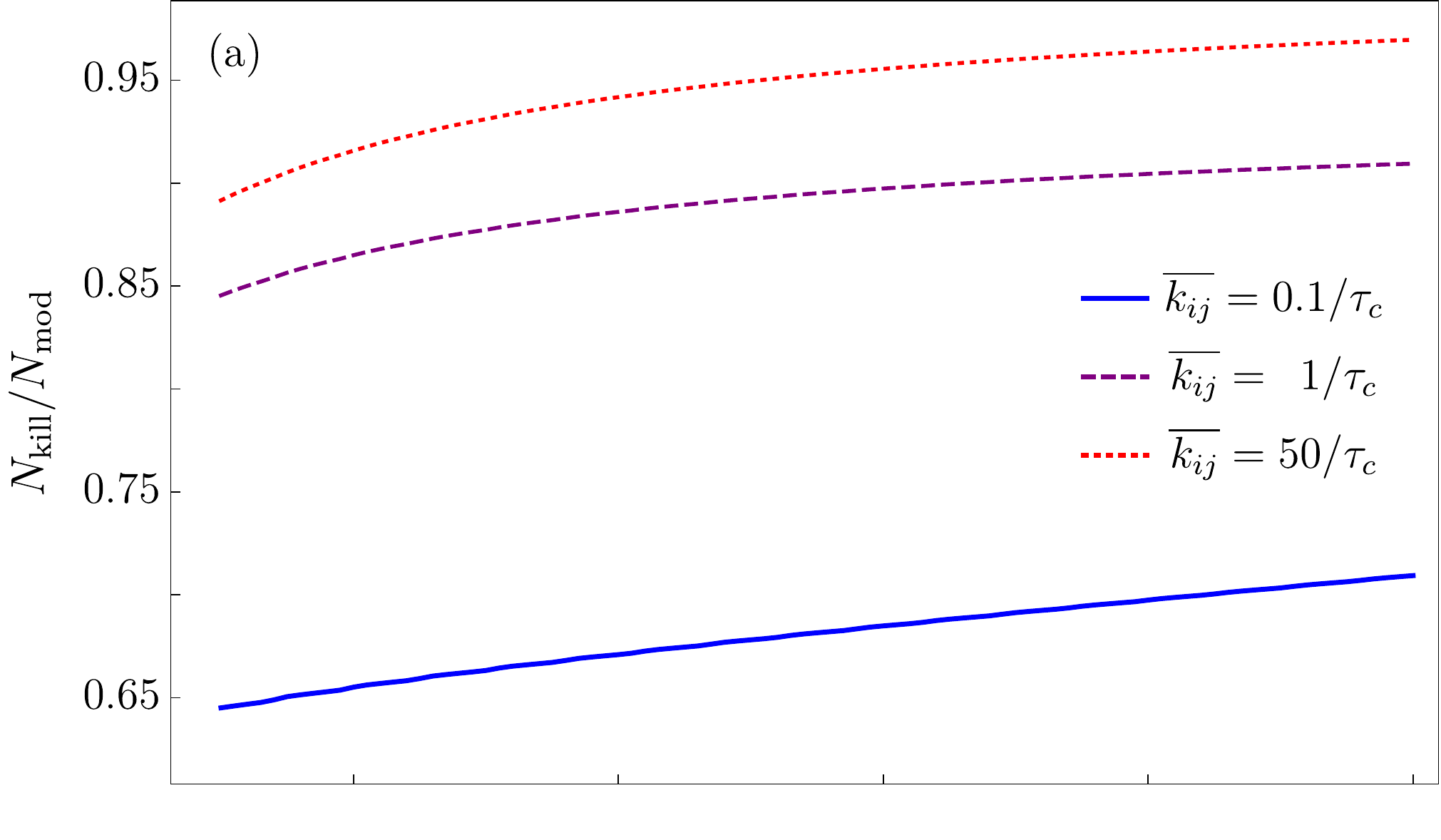}
\includegraphics[width=0.45\textwidth]{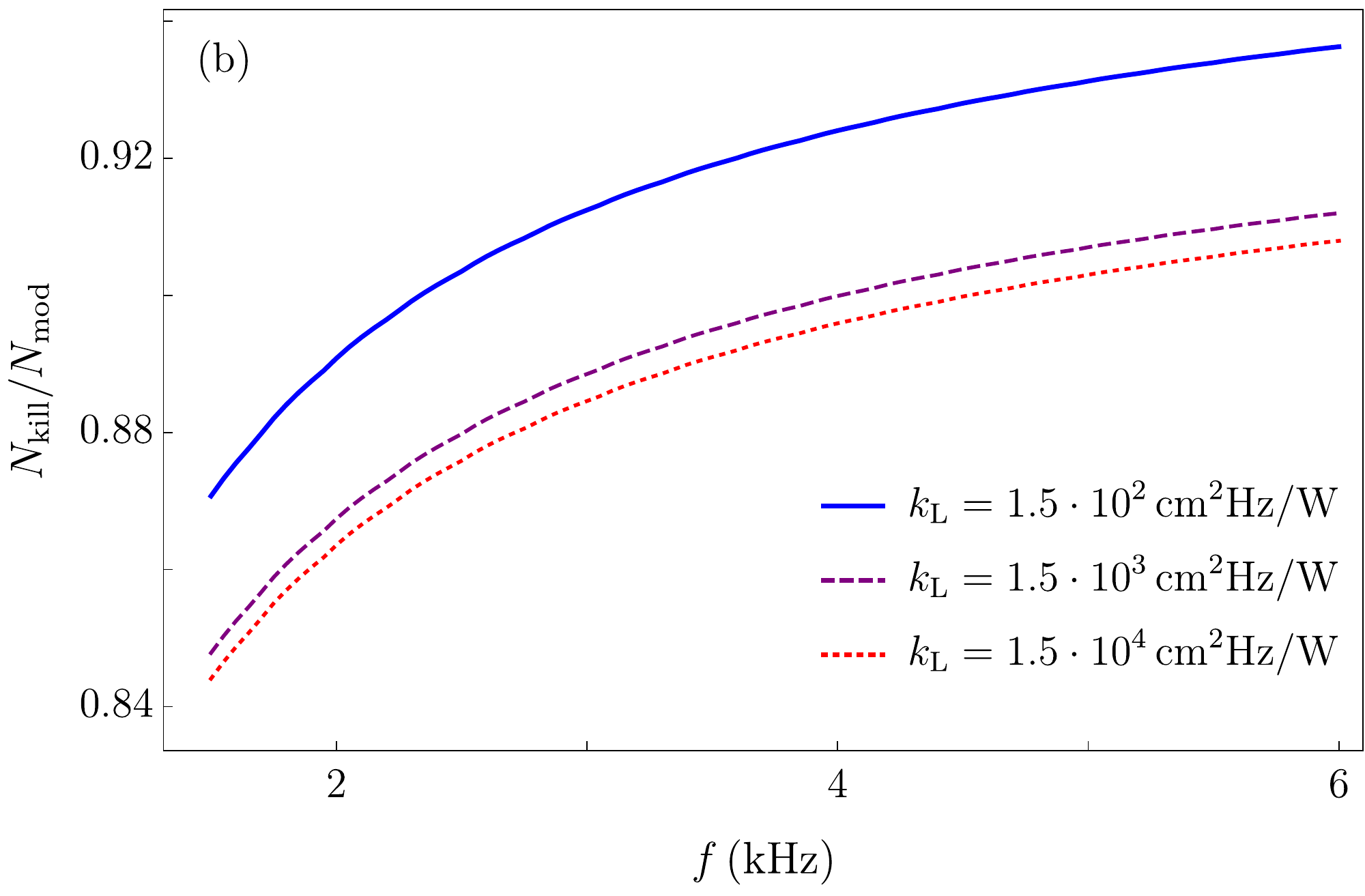}
\caption{\label{fig:rbcss} (a) The ratio between the molecules left in the trap with ($N_\text{kill}$) and without ($N_\text{mod}$) the additional light beam after waiting time of $400\,$ms as a function of the chopping rate $f$ for $^{87}$Rb$^{133}$Cs molecules assuming varying mean internal state coupling strength $\overline{k_{i,j}}$. (b)~Same, but for $\overline{k_{i,j}}=1/\tau_c$ and varying light absorption rates $k_L$. Here we assume $\tau_c=253\,\mu$s.}
%\caption{\label{fig:rbcss} (a) The ratio between the molecules left in the trap with ($N_\text{kill}$) and without ($N_\text{mod}$) the additional light beam after waiting time of $400\,$ms as a function of the chopping rate $f$ for RbCs molecules assuming varying mean internal state coupling strength $\overline{k_{i,j}}=0.1/\tau_c$ (blue solid), $1/\tau_c$ (purple dashed), and $50/\tau_c$ (red dotted). (b)~Same, but for $\overline{k_{i,j}}=1/\tau_c$ and varying light absorption rates $k_L=1.5\cdot 10^2\,$W$^{-1}$cm$^2$Hz (blue solid), $1.5\cdot 10^3$ (purple dashed), and $15\cdot 10^3$ (red dotted). Here we assume $\tau_c=253\,\mu$s.}
\end{figure}

In stark contrast, recent experiments with nonreactive molecules, namely fermionic $^{23}$Na$^{40}$K~\cite{Bause2021} as well as bosonic $^{23}$Na$^{87}$Rb and $^{23}$Na$^{39}$K~\cite{Gersema2021} reported disagreement with the semiclassical complex lifetime estimates. The results could only be made consistent with theory by assuming much longer complex lifetimes. For example, the metastable dimer composed of bosonic $^{23}$Na$^{39}$K would need to live for at least $0.3\,$ms, whereas the model prediction is about 50 times lower~\cite{ChristiansenPRA19}. As our estimates of the rate of nuclear spin-changing processes for these molecules lie in the range of kHz, they can provide a possible explanation of the observed disagreement. The impact of internal state couplings for the case of $^{23}$Na$^{40}$K with all parameters chosen to be similar to the Munich experiment~\cite{Bause2021} is showcased in Fig.~\ref{fig:nak}. We plot the ratio between the number of molecules left in the trap after waiting time of $250\,$ms with and without the kill beam with random couplings $k_{i,j}$ among all possible hyperfine levels for three different mean values of $k_{i,j}$ with respect to the bare decay time $\tau_c=18\,\mu$s. In agreement with the experimental results, we observe that as the couplings become stronger, the ratio approaches unity, and it becomes increasingly difficult to mitigate losses by chopping the trap and to measure the sensitivity of losses to the chopping rate.

However, the experiment performed with $^{87}$Rb$^{133}$Cs molecules~\cite{GregoryNC19,GregoryPRL20} reported vastly different dynamics. Strong dependence of losses on the chopping frequency has been demonstrated, leading to the estimate $\tau_c=0.53\,$ms, while the theory predicts $\tau_c=0.253\,$ms. For such a heavy system, the internal state couplings are presumably large, and it seems surprising that the semiclassical model is so successful. This can be explained within our model by taking into account the interplay with the light absorption timescales. Again, larger $k_{i,j}$ leads to lesser sensitivity to the presence of the kill beam as demonstrated in Fig.~\ref{fig:rbcss}(a). As the photoexcitation rate effectively increases, the complex absorbs photons before it can change the spin state. As a result, the internal states are less populated, and their impact is lowered, decreasing the $N_\text{kill}/N_\text{mod}$ ratio between the modulated trap and the one with the kill beam turned on, as shown in Fig.~\ref{fig:rbcss}(b). Then the observed lifetime can again be explained neglecting nuclear couplings. Note that for RbCs the estimated $k_L$ is much higher than for NaK~\cite{ChristiansenPRA19,GregoryPRL20}, in agreement with our conclusions.

{\it Atom-molecule collisions.} Seemingly larger disagreement between semiclassical model prediction and experimental measurement of the lifetime (by orders of magnitude) was recently reported for three-atom complexes formed in ultracold $^{40}$K$^{87}$Rb+$^{87}$Rb mixtures~\cite{NicholsPRX22}. Here couplings between hyperfine manifolds can also be expected to play an important role~\cite{FryeNJP21}. To verify this hypothesis, we calculate the interaction-induced variation of the leading hyperfine interaction between unpaired electronic spin and nuclear spins in the $^{40}$K$^{87}$Rb$_2$ complex~\cite{SupMat} (the interaction absent in the studied closed-shell four-atom complexes). We find that the decrease of the hyperfine coupling constant at the incoming open-shell Rb atom exceeds one GHz, while the coupling constants for the nuclear spins of the Rb and K atoms within colliding closed-shell molecule with incoming unpaired electronic spin emerges to over one GHz and hundreds of MHz at small atom-molecules distances, respectively. Predicted intermolecular hyperfine couplings are large enough to couple different hyperfine manifolds and can be responsible for the observed decay time enhancement in three-atom complexes.

{\it Experimental proposal.} Our predictions can be verified in experiments by comparing results involving ultracold molecules of elements having isotopes with and without nuclear spin, such as Sr, Yb, Cr, and O. We suggest that recently produced ultracold ground-state strontium dimers~\cite{Leung2021} are a suitable system for further studies and verification of the importance of the nuclear spin--spin and quadrupole couplings for the dynamics of intermediate complexes. Depending on the isotope, the Bose-Bose dimers do not have a hyperfine structure, while Fermi-Fermi dimers have spin--spin and nuclear quadrupole couplings. Because the dimer-dimer collisions are chemically reactive due to trimer formation, we propose employing Sr+Sr$_2$ mixtures to probe losses and lifetimes of intermediate three-atom complexes. For example, $^{88}$Sr$_2$+$^{88}$Sr has no hyperfine structure, $^{88}$Sr$_2$+$^{87}$Sr involves the nuclear quadruple coupling on the atom and its interaction-induced variation, $^{87}$Sr$_2$+$^{88}$Sr presents the nuclear quadruple, spin--spin, and spin--rotation couplings on the molecule and their interaction-induced variations, and, finally, $^{87}$Sr$_2$+$^{87}$Sr have hyperfine couplings scheme as in collisions of the $^1\Sigma$-state molecules. Observation of the strong dependence of the two-body losses on the isotope and the presence of hyperfine couplings confronted with theoretical models could confirm our predictions. Additionally, mixtures containing heteronuclear homoatomic dimers could provide more opportunities for testing chemical reactivity~\cite{TomzaPRL15}. Other possible experimental tests include the rotational excitation of molecules that directly actives nuclear quadrupole coupling and studies of chemically distinct molecules, such as deeply bound AlF ones~\cite{HofsassNJP21}.

{\it Conclusions.}
We have studied the internal structure of tetratomic complexes produced in bimolecular collisions of ultracold alkali-metal dimers and found significant couplings between nuclear states via the nuclear quadrupole and spin--spin interactions. Taking this into account in constructing a simplified model of reaction kinetics, we have been able to study the interplay between the bare complex lifetime, its internal dynamics, and the photoexcitation rate, resolving recent experimental findings. We have argued that similar mechanism due to a large variation of the electronic spin--nuclear spin coupling can be responsible for the unexpectedly large lifetime of the KRb-Rb complex. We have suggested possible experiments that would allow more insight into nuclear spin dynamics of ultracold complex states. It would now be interesting to study the dynamics of such complexes fully quantum mechanically, going beyond the simple kinetic equations and including the possibility of creating long-lived quantum scars and analogs of localization, as well as the impact of external electric field which would further affect the couplings structure.

\begin{acknowledgments}
We thank Paul S. Julienne for insightful discussions. This work was supported by the National Science Centre, Poland (grant 2020/37/B/ST2/00486), the Foundation for Polish Science within the First Team programme cofinanced by the European Union under the European Regional Development Fund, and the PL-Grid Infrastructure.
\end{acknowledgments}

\bibliography{A2B2}

\end{document}